\documentclass[sigconf,screen,authorversion]{acmart}

\copyrightyear{2024}
\acmYear{2024}
\setcopyright{acmlicensed}
\acmConference[SA Technical Communications '24]{SIGGRAPH Asia 2024 Technical Communications}{December 3--6, 2024}{Tokyo, Japan}
\acmBooktitle{SIGGRAPH Asia 2024 Technical Communications (SA Technical Communications '24), December 3--6, 2024, Tokyo, Japan}
\acmDOI{10.1145/3681758.3698000}
\acmISBN{979-8-4007-1140-4/24/12}

\setcitestyle{square}

\usepackage{xcolor}
\usepackage{wrapfig}
\newcommand{\set}[1]{\mathbf{#1}}
\newcommand{\eg}{\textit{e.g.}}
\newcommand{\ie}{\textit{i.e.}}

\begin{document}

\title[A Practical Style Transfer Pipeline for 3D Animation: Insights from Production R\&D]{A Practical Style Transfer Pipeline for 3D Animation:\\Insights from Production R\&D}

\author{Hideki Todo}
\orcid{0000-0002-3010-3401}
\affiliation{%
    \institution{Takushoku University}
    \country{Japan}
}
\email{htodo@cs.takushoku-u.ac.jp}

\author{Yuki Koyama}
\orcid{0000-0002-3978-1444}
\affiliation{%
    \institution{Graphinica, Inc.}
    \country{Japan}
}
\email{koyama.yuki@graphinica.com}

\author{Kunihiro Sakai}
\orcid{0009-0003-4582-242X}
\affiliation{%
    \institution{Graphinica, Inc.}
    \country{Japan}
}
\email{sakai.kunihiro@graphinica.com}

\author{Akihiro Komiya}
\orcid{0009-0003-4939-6667}
\affiliation{%
    \institution{Graphinica, Inc.}
    \country{Japan}
}
\email{komiya.akihiro@graphinica.com}

\author{Jun Kato}
\orcid{0000-0003-4832-8024}
\affiliation{%
    \institution{Arch Inc.}
    \country{Japan}
}
\email{jun@archinc.jp}

\begin{teaserfigure}
    \centering
    \includegraphics[width=\linewidth]{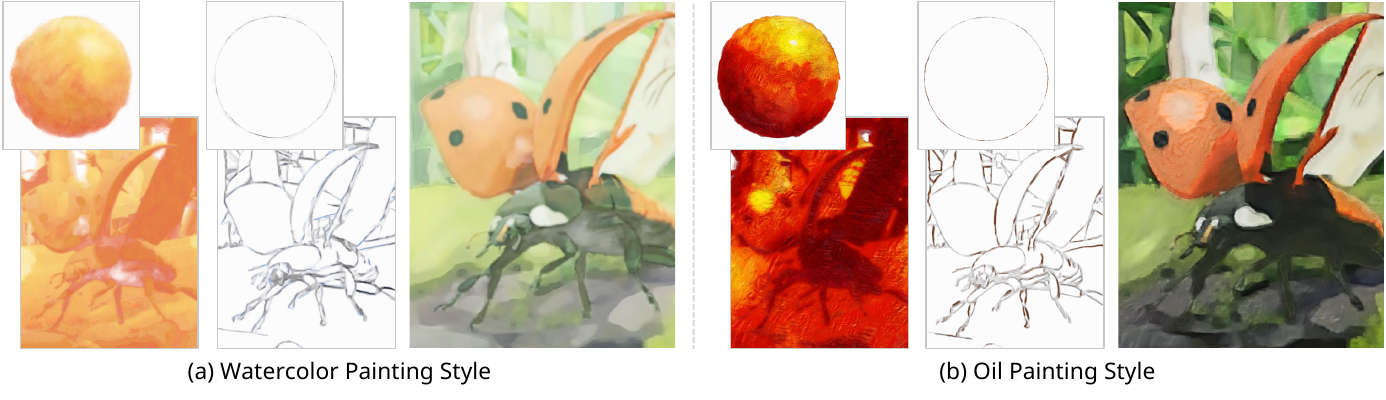}
    \caption{
        Style variations generated by our style transfer pipeline: (a) watercolor and (b) oil painting styles.
        We use style transfer to generate intermediate results (base touch and outline layers), which are then used to produce high-quality final compositions.
    }
    \label{fig:teaser}
    \vspace{2mm}
\end{teaserfigure}

\begin{abstract}
    % !TEX root = ../main_long.tex

Our animation studio has developed a practical \emph{style transfer} pipeline for creating stylized 3D animation, which is suitable for complex real-world production.
This paper presents the insights from our development process, where we explored various options to balance quality, artist control, and workload, leading to several key decisions.
For example, we chose patch-based texture synthesis over machine learning for better control and to avoid training data issues.
We also addressed specifying style exemplars, managing multiple colors within a scene, controlling outlines and shadows, and reducing temporal noise.
These insights were used to further refine our pipeline, ultimately enabling us to produce an experimental short film showcasing various styles.

\end{abstract}

\begin{CCSXML}
    <ccs2012>
    <concept>
    <concept_id>10010147.10010371.10010372.10010375</concept_id>
    <concept_desc>Computing methodologies~Non-photorealistic rendering</concept_desc>
    <concept_significance>500</concept_significance>
    </concept>
    </ccs2012>
\end{CCSXML}

\ccsdesc[500]{Computing methodologies~Non-photorealistic rendering}

\keywords{Production R\&D, Style Transfer, Non-Photorealistic Rendering}

\maketitle

% !TEX root = ../main_long.tex

\begin{figure*}
    \centering
    \includegraphics[width=\linewidth]{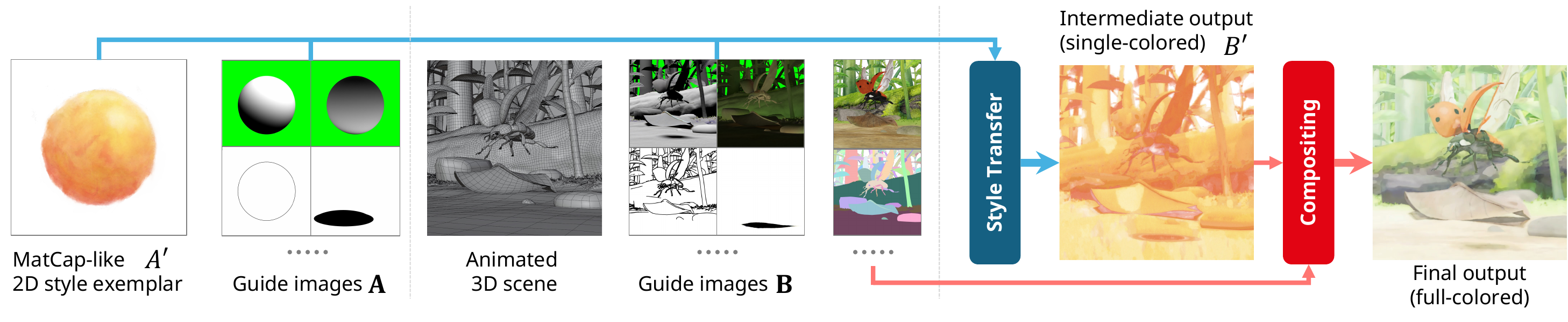}
    \caption{
        The overview of the proposed style transfer pipeline.
    }
    \label{fig:overview}
\end{figure*}

\section{Introduction}
\label{sec:introduction}

\paragraph{Goals}
In the pursuit of practical non-photorealistic rendering (NPR) techniques suitable for actual production, our animation studio has conducted research and development (R\&D) on a production pipeline for creating stylized 3D animation (see \autoref{fig:teaser} for results).
To support artistic creativity, we set the following goals:
\begin{itemize}
    \item \textbf{Expression augmentation}: Enhancing artistic expression of 3D animation, particularly for animating the unique and diverse touches of 2D hand-drawn art (\eg, watercolor and oil paintings),
    \item \textbf{Complex scene suitability}: Handling intricate production scenes with many objects and materials, bridging the gap between academic research and real-world production,
    \item \textbf{Artist control}: Providing high artist control over the process, such as generating intermediate outputs that are usable in the compositing stage, rather than generating final outputs by a fully automated black-box process, and
    \item \textbf{Practical workload}: Balancing quality and effort, addressing real-world production complexities.
\end{itemize}

\paragraph{Our Approach and Challenge}
In this project, we decided to take the \emph{style transfer} approach \cite{Hertzmann:SIGGRAPH:01} to support diverse styles.
Despite its long research history, this approach has not yet been widely used in actual production, making it challenging to design and implement a practical production pipeline.
For example, we needed to decide which algorithm to use, how to specify target styles, how to handle complex scenes, and how to reduce artificial noise.

\paragraph{Contributions}
This paper reports on the insights gained during our R\&D process, including key decision-making discussions and findings from experiments.
We explored various options with the above goals in mind.
We believe our insights contribute to the graphics community by assisting R\&D in other animation studios as well as informing academic researchers about production considerations.

\section{A Practical Style Transfer Pipeline}

\autoref{fig:overview} shows the style transfer pipeline we finally developed.
In addition to the target 3D scene, the style transfer stage takes a \emph{style exemplar} 2D image as input.
The output of the style transfer stage is single-colored at this point and is used at the compositing stage to create full-colored stylized animation.
The style transfer stage follows the \emph{Image Analogies} formulation \cite{Hertzmann:SIGGRAPH:01}:
given the style exemplar $A'$ and guide images $\set{A} = \{ A^{1}, \ldots, A^{n} \}$ (for the style exemplar) and $\set{B} = \{ B^{1}, \ldots, B^{n} \}$ (for the target scene), where $n$ is the number of guide images, generate a stylized target image $B'$ such that the relationship
\begin{align}
    \set{A}:A' :: \set{B}:B'
\end{align}
holds.
For guide images, we combined various information extracted from the 3D scene (\eg, diffuse, normal, and outlines), depending on the scenes.

\subsection{Style Transfer Algorithm}
Style transfer has a long history with two main approaches:
classical \emph{patch-based texture synthesis} (\eg, \cite{Hertzmann:SIGGRAPH:01}) and recent \emph{neural} methods (\eg, \cite{Gatys:CVPR:2016,Texler:SIGGRAPH:20}).
We tested both and found that neural methods often lose fine texture details (crucial for us), and their training is not sufficiently stable and predictable (sensitive to hyperparameters and random seeds).
In contrast, the patch-based approach provided stable outputs with fine details, leading us to choose this approach.

\subsection{MatCap-Like Style Exemplar}
We considered two options for style exemplar formats (\ie, how to specify the target style):
\begin{itemize}
    \item \textbf{The keyframing approach}: drawing stylized touches on several keyframes of the target animation (\eg, \cite{Texler:SIGGRAPH:20}) and
    \item \textbf{The MatCap-like approach}: drawing stylized touches on a reference sphere scene (\eg, \cite{Fivser:SIGGRAPH:16}).
\end{itemize}
Note that we use the term ``MatCap-like'' (rather than ``MatCap'') because it differs in certain respects from traditional MatCap \cite{Todo:CGI:13,Koyama:SA:2021}.
In MatCap, styles are strictly defined by drawing touches and shading within the contour of a sphere.
Our MatCap-like approach also defines styles by drawing on a sphere, but it allows for variations near the contour, accommodating characteristics of 2D analog materials, such as incomplete or overextended strokes.
Additionally, it permits the definition of styles for shadows and backgrounds outside the sphere, further distinguishing it from the conventional MatCap method.

We chose the simple yet flexible MatCap-like format, allowing artists to design stylized touches as general materials easily applicable to individual scenes.
Compared to keyframing, the asset preparation cost is significantly lower;
for example, if we draw three keyframes for each cut and we have 100 cuts, the keyframing approach requires drawing 300 style exemplars;
on the other hand, the MatCap-like approach requires drawing only one style exemplar in the same situation.
The transfer quality of the MatCap-like approach was sufficiently high and robust for our usage.

\subsection{Handling of Multiple Colors}
\begin{figure}
    \centering
    \includegraphics[width=\linewidth]{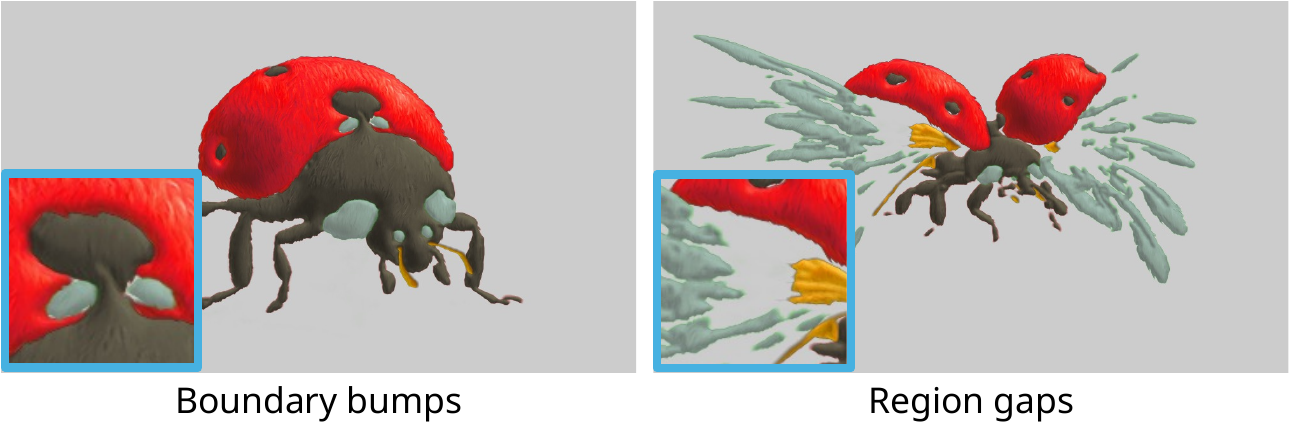}
    \caption{
        Artifacts of the tested na\"{i}ve approach (\ie, perform style transfer for each color and then combine).
    }
    \label{fig:naive-multi-color}
\end{figure}

With the MatCap-like style specification, applying style transfer to scenes with multiple colors is challenging.
We tested a na\"{i}ve approach where each color is transferred individually and then combined together;
however, this resulted in artifacts such as boundary bumps and region gaps (see \autoref{fig:naive-multi-color} and the supplemental video).

This led us to a more robust solution:
transferring only the touch texture of a single target medium to the entire scene, resulting in a single-colored output, and then performing coloring in the compositing stage.
This solution was robust even with complex scenes and also simplified the pipeline.

\subsection{Separation of Outlines and Shadows}
Our pipeline runs style transfer for the outline and base touch layers separately (see \autoref{fig:teaser}).
This approach has two major advantages.
First, it allows for different transfer settings for each layer, making artist control easier.
Second, it facilitates precise color control during the compositing stage (\eg, the composite artist can apply different color filters for each layer).
In the same way, we also found it useful to separate shadows for easier adjustment.

Note that it was our key finding that stylized outlines can be synthesized surprisingly well using the patch-based algorithm, retaining hand-drawn details.
To the best of our knowledge, this insight has not been widely recognized.

\subsection{Temporal Noise Reduction}
We found that the na\"{i}ve approach of synthesizing each frame independently (na\"{i}ve frame-by-frame synthesis) suffered from severe temporal noise (see the supplemental video).
To reduce the temporal noise, we tested an existing \emph{temporal guide} method based on motion field \cite{Jamrivska:SIGGRPAH:19}.
However, we found it impractical for us due to asset preparation costs and tracking errors caused by occlusion.

\begin{figure}
    \centering
    \includegraphics[width=\linewidth]{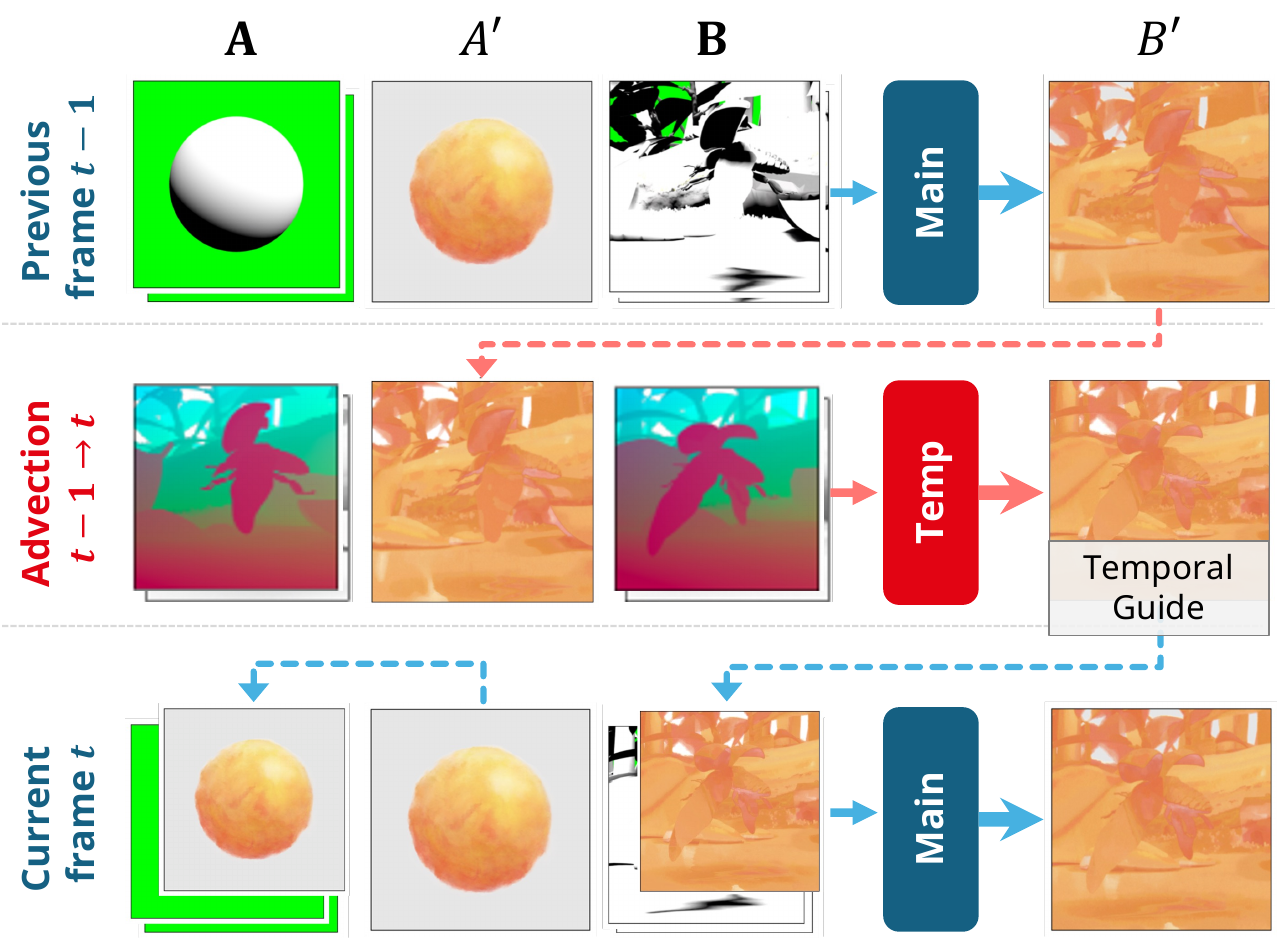}
    \caption{
        The concept of temporal noise reduction using an additional style transfer step for advection.
    }
    \label{fig:temporal-guide}
\end{figure}

Instead, we propose a simple temporal guide using an \emph{additional} style transfer step for performing advection of stylized touches.
See \autoref{fig:temporal-guide} for the overview of how our temporal guide is computed.
It involves the following style transfer steps.
\begin{itemize}
    \item Main style transfer step at the previous frame $t-1$:
          \begin{align}
              \set{A}:A' :: \set{B}_{t-1}:B'_{t-1} \nonumber
          \end{align}
    \item Additional style transfer step for advection from the previous frame $t-1$ to the current frame $t$:
          \begin{align}
              \set{A}_\text{adv}:A'_\text{adv} :: \set{B}_\text{adv}:B'_\text{adv} \nonumber
          \end{align}
    \item Main style transfer step at the current frame $t$:
          \begin{align}
              \set{A}:A' :: \set{B}_{t}:B'_{t} \nonumber
          \end{align}
\end{itemize}

We consider the transfer result of the previous frame, $B'_{t-1}$, as the style exemplar for the advection step, $A'_\text{adv}$, and apply the style transfer to create the advected result, $B'_\text{adv}$.
To ensure alignment across adjacent frames, we primarily use the \emph{world position} guide (\ie, an RGB image where RGB corresponds to XYZ with normalization) for the guide images $\set{A}_\text{adv}$ and $\set{B}_\text{adv}$.
This advected result $B'_\text{adv}$ is then used as an additional guide (called a temporal guide) in the main transfer step at the current frame $t$ (\ie, $B'_\text{adv}$ is added to the definition of the target guide images at the current frame, $\set{B}_{t}$) to stabilize the touch placements.
For the style guide images $\set{A}$, the style exemplar $A'$ is added to be coupled with the added $B'_\text{adv}$.
Finally, our main style transfer step with the temporal guide is defined as:
\begin{align}
    \underbrace{\{A^{1}, \ldots, A^{n}, A'\}}_{\set{A}}:A' :: \underbrace{\{ B^{1}_{t}, \ldots, B^{n}_{t}, B'_\text{adv} \}}_{\set{B}_{t}}:B'_{t}.
\end{align}
Note that $\set{A}$ and $A'$ remain unchanged through all the frames, while $\set{B}_{t}$ are updated sequentially for each frame.

This approach significantly reduced temporal noise (see the supplemental video).
As an additional benefit, we found it useful for reducing boundary artifacts found in the baseline setting (ours without temporal guides) and in a neural approach \cite{Texler:SIGGRAPH:20} (see the supplemental video).

\section{Results}

To test our pipeline, we prepared an example test scene with a production-level complexity, including an animated character (a ladybug), a background scene with many objects, and a non-static camera motion.
We also asked 2D artists to create four different style exemplars: watercolor, oil, hatching, and pastel styles.

\begin{figure*}
    \centering
    \includegraphics[width=\linewidth]{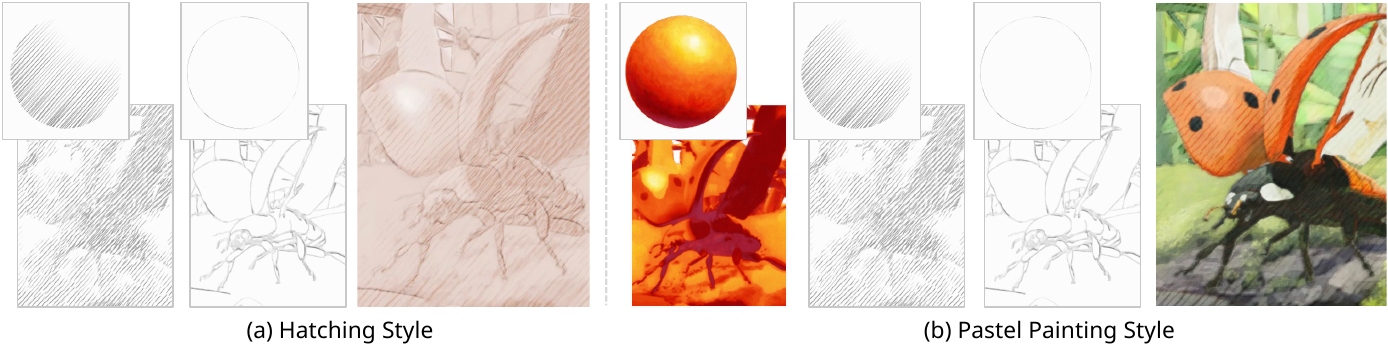}
    \caption{
        Additional style variations generated by our style transfer pipeline: (a) hatching and (b) pastel painting styles.
    }
    \label{fig:another_results}
\end{figure*}

\autoref{fig:teaser} and \autoref{fig:another_results} show the results.
Note that the result of the pastel style was created by not only using the style transfer result of the base pastel touch but also adding the details with the outputs of the hatching style transfer;
this approach of mixing multiple styles was enabled by the flexibility of our pipeline (\ie, providing intermediate results rather than final results) and facilitated the creative exploration of artistic looks.

\section{Short Film Production}

\begin{figure*}
    \centering
    \includegraphics[width=\linewidth]{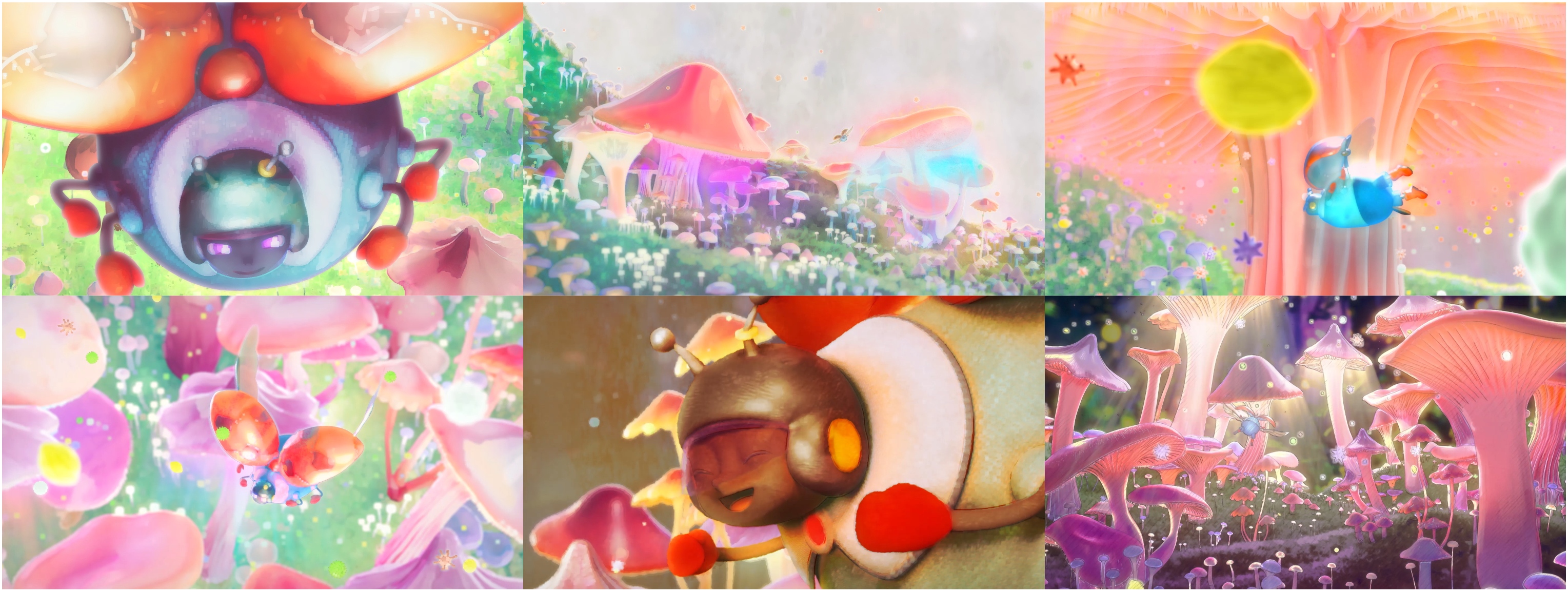}
    \caption{
        Production still images from our experimental short film, highlighting various styles effectively communicating the film's atmosphere.
        \copyright{} ARCH $\cdot$ Graphinica $\cdot$ Salamander Pictures
    }
    \label{fig:film}
\end{figure*}

With the success of the above style transfer pipeline applied to the example scene (\autoref{fig:teaser} and \autoref{fig:another_results}), we decided to create an experimental short film artwork.
Based on the insights we gained, we further refined the pipeline for real production in collaboration with artists.
The completed animation is available on \href{https://www.youtube.com/watch?v=EJ_vwZbOFCs}{YouTube}.

The short film features various hand-drawn styles (see \autoref{fig:film}), such as watercolor painting and pencil line drawing,
which are difficult to achieve without our pipeline (\textbf{expression augmentation}).
This film includes a cut scene that combines multiple styles together, which was enabled by providing useful intermediate style transfer results for the compositing stage (\textbf{artist control}).
The MatCap-like style exemplar format was compact and easy to manage assets, yet effectively worked for production-level complex scenes (\textbf{complex scene suitability}).
Also, this format allows us to make trials and errors of style transfer in the early stage of the entire project and finalize the look and feel in the early stage consequently, which was highly helpful for avoiding an additional significant workload of trials and errors for each scene (\textbf{practical workload}).

\begin{acks}
    We thank Kunio Moteki, Makoto Nakajima, and Remy Torre for their contributions to R\&D.
    We also appreciate Michael Arias and everyone involved in the short film production from ARCH, Graphinica, and Salamander Pictures.
\end{acks}

\bibliographystyle{ACM-Reference-Format}
\bibliography{main}


\begin{thebibliography}{7}

%%% the Web address.

\ifx \showCODEN    \undefined \def \showCODEN     #1{\unskip}     \fi
\ifx \showDOI      \undefined \def \showDOI       #1{#1}\fi
\ifx \showISBNx    \undefined \def \showISBNx     #1{\unskip}     \fi
\ifx \showISBNxiii \undefined \def \showISBNxiii  #1{\unskip}     \fi
\ifx \showISSN     \undefined \def \showISSN      #1{\unskip}     \fi
\ifx \showLCCN     \undefined \def \showLCCN      #1{\unskip}     \fi
\ifx \shownote     \undefined \def \shownote      #1{#1}          \fi
\ifx \showarticletitle \undefined \def \showarticletitle #1{#1}   \fi
\ifx \showURL      \undefined \def \showURL       {\relax}        \fi
% invisible to TeX
\providecommand\bibfield[2]{#2}
\providecommand\bibinfo[2]{#2}
\providecommand\natexlab[1]{#1}
\providecommand\showeprint[2][]{arXiv:#2}

\bibitem[Fi\v{s}er et~al\mbox{.}(2016)]%
        {Fivser:SIGGRAPH:16}
\bibfield{author}{\bibinfo{person}{Jakub Fi\v{s}er},
  \bibinfo{person}{Ond\v{r}ej Jamri\v{s}ka}, \bibinfo{person}{Michal
  Luk\'{a}\v{c}}, \bibinfo{person}{Eli Shechtman}, \bibinfo{person}{Paul
  Asente}, \bibinfo{person}{Jingwan Lu}, {and} \bibinfo{person}{Daniel
  S\'{y}kora}.} \bibinfo{year}{2016}\natexlab{}.
\newblock \showarticletitle{StyLit: illumination-guided example-based
  stylization of 3D renderings}.
\newblock \bibinfo{journal}{\emph{ACM Trans. Graph.}} \bibinfo{volume}{35},
  \bibinfo{number}{4}, Article \bibinfo{articleno}{92} (\bibinfo{year}{2016}),
  \bibinfo{numpages}{11}~pages.
\newblock
\urldef\tempurl%
\url{https://doi.org/10.1145/2897824.2925948}
\showDOI{\tempurl}


\bibitem[Gatys et~al\mbox{.}(2016)]%
        {Gatys:CVPR:2016}
\bibfield{author}{\bibinfo{person}{Leon~A. Gatys},
  \bibinfo{person}{Alexander~S. Ecker}, {and} \bibinfo{person}{Matthias
  Bethge}.} \bibinfo{year}{2016}\natexlab{}.
\newblock \showarticletitle{Image Style Transfer Using Convolutional Neural
  Networks}. In \bibinfo{booktitle}{\emph{Proc. CVPR '16}}.
  \bibinfo{pages}{2414--2423}.
\newblock
\urldef\tempurl%
\url{https://doi.org/10.1109/CVPR.2016.265}
\showDOI{\tempurl}


\bibitem[Hertzmann et~al\mbox{.}(2001)]%
        {Hertzmann:SIGGRAPH:01}
\bibfield{author}{\bibinfo{person}{Aaron Hertzmann},
  \bibinfo{person}{Charles~E. Jacobs}, \bibinfo{person}{Nuria Oliver},
  \bibinfo{person}{Brian Curless}, {and} \bibinfo{person}{David~H. Salesin}.}
  \bibinfo{year}{2001}\natexlab{}.
\newblock \showarticletitle{Image analogies}. In
  \bibinfo{booktitle}{\emph{Proc. SIGGRAPH '01}}. \bibinfo{pages}{327--340}.
\newblock
\urldef\tempurl%
\url{https://doi.org/10.1145/383259.383295}
\showDOI{\tempurl}


\bibitem[Jamri\v{s}ka et~al\mbox{.}(2019)]%
        {Jamrivska:SIGGRPAH:19}
\bibfield{author}{\bibinfo{person}{Ond\v{r}ej Jamri\v{s}ka},
  \bibinfo{person}{\v{S}\'{a}rka Sochorov\'{a}}, \bibinfo{person}{Ond\v{r}ej
  Texler}, \bibinfo{person}{Michal Luk\'{a}\v{c}}, \bibinfo{person}{Jakub
  Fi\v{s}er}, \bibinfo{person}{Jingwan Lu}, \bibinfo{person}{Eli Shechtman},
  {and} \bibinfo{person}{Daniel S\'{y}kora}.} \bibinfo{year}{2019}\natexlab{}.
\newblock \showarticletitle{Stylizing video by example}.
\newblock \bibinfo{journal}{\emph{ACM Trans. Graph.}} \bibinfo{volume}{38},
  \bibinfo{number}{4}, Article \bibinfo{articleno}{107} (\bibinfo{year}{2019}),
  \bibinfo{numpages}{11}~pages.
\newblock
\urldef\tempurl%
\url{https://doi.org/10.1145/3306346.3323006}
\showDOI{\tempurl}


\bibitem[Koyama et~al\mbox{.}(2021)]%
        {Koyama:SA:2021}
\bibfield{author}{\bibinfo{person}{Yuki Koyama}, \bibinfo{person}{Takeshi
  Tsuruta}, \bibinfo{person}{Heisuke Saito}, \bibinfo{person}{Daisuke
  Takizawa}, {and} \bibinfo{person}{Hiroshi Moriguchi}.}
  \bibinfo{year}{2021}\natexlab{}.
\newblock \showarticletitle{A Procedural MatCap System for Cel-Shaded Japanese
  Animation Production}. In \bibinfo{booktitle}{\emph{SIGGRAPH Asia 2021
  Posters}}. \bibinfo{pages}{34:1--34:2}.
\newblock
\urldef\tempurl%
\url{https://doi.org/10.1145/3476124.3488620}
\showDOI{\tempurl}


\bibitem[Texler et~al\mbox{.}(2020)]%
        {Texler:SIGGRAPH:20}
\bibfield{author}{\bibinfo{person}{Ond\v{r}ej Texler}, \bibinfo{person}{David
  Futschik}, \bibinfo{person}{Michal ku\v{c}era}, \bibinfo{person}{Ond\v{r}ej
  jamri\v{s}ka}, \bibinfo{person}{\v{S}\'{a}rka Sochorov\'{a}},
  \bibinfo{person}{Menclei Chai}, \bibinfo{person}{Sergey Tulyakov}, {and}
  \bibinfo{person}{Daniel S\'{y}kora}.} \bibinfo{year}{2020}\natexlab{}.
\newblock \showarticletitle{Interactive video stylization using few-shot
  patch-based training}.
\newblock \bibinfo{journal}{\emph{ACM Trans. Graph.}} \bibinfo{volume}{39},
  \bibinfo{number}{4}, Article \bibinfo{articleno}{73} (\bibinfo{year}{2020}),
  \bibinfo{numpages}{11}~pages.
\newblock
\urldef\tempurl%
\url{https://doi.org/10.1145/3386569.3392453}
\showDOI{\tempurl}


\bibitem[Todo et~al\mbox{.}(2013)]%
        {Todo:CGI:13}
\bibfield{author}{\bibinfo{person}{Hideki Todo}, \bibinfo{person}{Ken Anjyo},
  {and} \bibinfo{person}{Shun'Ichi Yokoyama}.} \bibinfo{year}{2013}\natexlab{}.
\newblock \showarticletitle{Lit-Sphere extension for artistic rendering}.
\newblock \bibinfo{journal}{\emph{Vis. Comput.}} \bibinfo{volume}{29},
  \bibinfo{number}{6--8} (\bibinfo{year}{2013}), \bibinfo{pages}{473--480}.
\newblock
\urldef\tempurl%
\url{https://doi.org/10.1007/s00371-013-0811-7}
\showDOI{\tempurl}


\end{thebibliography}

\end{document}